\documentclass{elsart}

\usepackage{graphicx}
\usepackage{amsmath}
\usepackage{amsfonts}
\usepackage{amssymb}

\newlength{\zero}
\begin{document}

\begin{frontmatter}

\title{AMR simulations of the low $T/|W|$ bar-mode instability of neutron stars}

\author{Pablo Cerd\'a-Dur\'an, Vicent Quilis, and Jos\'e A. Font}

\address{Departamento de Astronom\'{\i}a y Astrof\'{\i}sica,
  Universidad de Valencia, Dr. Moliner 50, 46100 Burjassot
  (Valencia), Spain}

\begin{abstract}
It has been recently argued through numerical work that rotating stars with a 
high degree of differential rotation are dynamically unstable against bar-mode 
deformation, even for values of the ratio of rotational kinetic energy to 
gravitational potential energy as low as $\cal{O}$$(0.01)$. This may have 
implications for gravitational wave astronomy in high-frequency sources such 
as core collapse supernovae. In this paper we present high-resolution simulations, 
performed with an adaptive mesh refinement hydrodynamics code, of such low 
$T/|W|$ bar-mode instability. The complex morphological features involved in 
the nonlinear dynamics of the instability are revealed in our simulations, which 
show that the excitation of Kelvin-Helmholtz-like fluid modes outside the 
corotation radius of the star leads to the saturation of the bar-mode deformation.
While the overall trends reported in an earlier investigation are confirmed by 
our work, we also find that numerical resolution plays an important role during 
the long-term, nonlinear behaviour of the instability, which has implications on 
the dynamics of rotating stars and on the attainable amplitudes of the associated 
gravitational wave signals.
\end{abstract}

\end{frontmatter}

\section{Introduction}

Neutron stars following a core collapse supernova are rotating at birth and
can be subject to various nonaxisymmetric instabilities 
(see e.g. \cite{stergioulas} for a review).
Among those, if the rotation rate is high enough so that the 
ratio of rotational kinetic energy $T$ to gravitational potential energy $W$, 
$\beta\equiv T/|W|$, exceeds the critical value $\beta_{\mathrm d}\sim 0.27$, 
inferred from studies with incompressible Maclaurin spheroids, the star is 
subject to a {\it dynamical} bar-mode ($l=m=2$ $f$-mode) instability driven 
by hydrodynamics and gravity. Its study is highly motivated nowadays as such 
an instability bears important implications in the prospects of detection of 
gravitational radiation from newly-born rapidly rotating neutron stars.

Simulations of the dynamical bar-mode instability are available in the
literature, both using simplified models based on equilibrium stellar 
configurations perturbed with suitable 
eigenfunctions~\cite{tohline85,houser94,new00,shibata00}, and more involved 
models for the core collapse scenario~\cite{rampp98,shibata05,saijo05,ott05},
and in either case both in Newtonian gravity and general relativity. Due to its 
superior simplicity the former approach has received much more attention, 
notwithstanding that the conclusions drawn from perturbed stellar models 
may not be straightforwardly extended to the collapse scenario. 

Newtonian simulations of triaxial instabilities following core collapse 
were first performed by~\cite{rampp98}. These showed that the bar-mode instability 
sets in when $\beta\gg 0.27$ and when the progenitor rotates rapidly and highly 
differentially. Such conditions are met when the (artificial) depletion of
internal energy to trigger the collapse is large enough to produce a very 
compact core for which a significant spun-up can be achieved. More recently, 
three-dimensional simulations of the core collapse of rotating polytropes in 
general relativity have been performed by~\cite{shibata05}. These authors studied 
the evolution of the bar-mode instability starting with axisymmetric core collapse 
initial models which reached values of $\beta\sim 0.27$ during the infall phase.
These simulations showed that the maximum value of $\beta$ achieved during 
collapse and bounce depends strongly on the velocity profile, the total mass 
of the initial core, and on the equation of state. In agreement with the findings 
from the Newtonian simulations of~\cite{rampp98}, the bar-mode instability sets 
in if the progenitor rotates rapidly ($0.01\le\beta\le 0.02$) and has a high 
degree of differential rotation. In addition, the artificial depletion of 
pressure and internal energy to trigger the collapse, leading to a compact 
core which subsequently spins up, also plays a key role in general relativity 
for a noticeable growth of the bar-mode instability.

Whether the requirements inferred from numerical simulations are at all met by the
collapse progenitors remains unclear. As shown by~\cite{spruit98} magnetic torques
can spin down the core of the progenitor, which leads to slowly rotating neutron 
stars at birth ($\sim 10-15$ms). The most recent, state-of-the-art computations 
of the evolution of massive stars, which include angular momentum redistribution 
by magnetic torques and spin estimates of neutron stars at 
birth~\cite{heger05,ott06}, lead to core collapse progenitors which do not seem 
to rotate fast enough to guarantee the unambiguous growth of the canonical bar-mode 
instability. Rapidly-rotating cores might be produced by an appropriate mixture of 
high progenitor mass ($M>25M_{\odot}$) and low metallicity (N.~Stergioulas, private
communication). In such case the progenitor could by-pass the Red Supergiant phase 
in which the differential rotation of the core produces a magnetic field by dynamo 
action which couples the core to the outer layers of the star, transporting angular 
momentum outwards and spinning down the core. According to~\cite{woosley06} about
1\% of all stars with $M>10M_{\odot}$ will produce rapidly-rotating cores.

On the other hand, Newtonian simulations of the bar-mode instability from perturbed 
equilibrium models of rotating stars have shown that $\beta_{\mathrm d}\sim 0.27$ 
independent of the stiffness of the equation of state 
provided the star is not strongly differentially rotating. The relativistic 
simulations of~\cite{shibata00} yielded a value of $\beta\sim 0.24-0.25$  
for the onset of the instability, while the dynamics of the process 
closely resembles that found in Newtonian theory, i.e.~unstable models with 
large enough $\beta$ develop spiral arms following the formation of bars, 
ejecting mass and redistributing the angular momentum. As the degree of 
differential rotation becomes higher Newtonian simulations have also 
shown that $\beta_{\mathrm d}$ can be as low as $0.14$~\cite{centrella01}.  
More recently~\cite{shibata02,shibata03} have reported that rotating stars 
with an {\it extreme} degree of differential rotation are dynamically unstable 
against bar-mode deformation even for values of $\beta$ of $\cal{O}$$(0.01)$. 

Given its recent discovery and its potential astrophysical implications for 
post-bounce core collapse dynamics and gravitational wave astronomy, we present
in this paper high resolution simulations of such low  $T/|W|$ bar-mode 
instabilities. This work is further motivated in the light of the few numerical 
simulations available in the literature. Our main goal is to revisit the 
simulations by~\cite{shibata02} on the low $T/|W|$ bar-mode instability, and 
particularly to check how sensitive the onset and development of the instability 
is to numerical issues such as grid resolution. To this aim we perform Newtonian 
hydrodynamical simulations of a subset of models analyzed by~\cite{shibata02} 
using an adaptive mesh refinement (AMR) code~\cite{quilis04} which allows us 
to perform such three dimensional simulations with the highest resolution ever 
used. Our simulations reveal the complex morphological features involved in the 
nonlinear dynamics of the instability, where the excitation of 
Kelvin-Helmholtz-like fluid modes influences the saturation of the bar-mode 
deformation. We advance that while the overall trends found by~\cite{shibata02} 
are confirmed by our work, the resolution employed in the simulations does play 
a key role for the long-term behaviour of the instability and for the nonlinear 
dynamics of rotating stars, which has implications on the attainable amplitudes 
of the associated gravitational wave signals. We note that we plan to upgrade the 
existing AMR code to account for the effects of magnetic fields in order to attempt 
the current study in a more realistic setup. The present work is a step towards 
that goal.

The paper is organized as follows: Section~\ref{eqs} gives a brief overview of
the equations to solve. Their solution is outlined in Section~\ref{numerics}
which also contains the bare details of the AMR code. The results of the simulations
are discussed in Section~\ref{results}. Finally Section~\ref{conclusions} presents
our conclusions.

\section{Mathematical framework}
\label{eqs}

The evolution of a self-gravitating ideal fluid in the Newtonian limit is described 
by the hydrodynamics equations and Poisson's equation:
\begin{equation}
\frac{\partial   \rho}{\partial  t}  +   \nabla\cdot(\rho{\bf v}) = 0
\label{hydro1}
\end{equation}
\begin{equation}
\frac{\partial  {\bf v}}{\partial t}  +   ({\bf v}\,\cdot\,
\nabla){\bf v} = - \frac{1}{\rho}\nabla p - \nabla \phi
\label{hydro2}
\end{equation}
\begin{equation}
\frac{\partial E}{\partial  t} +  \nabla\cdot[(E +  p) {\bf
v}] =   -  \rho {\bf v} \nabla \phi
\label{hydro3}
\end{equation}
\begin{equation}
\nabla^2\phi = 4\pi G \rho
\label{poisson1}
\end{equation}
where ${\bf  x}$, ${\bf v}=\frac{d{\bf x}}{dt}=  (v_x, v_y, v_z)$,
and $\phi(t,{\bf x})$ are, respectively, the Eulerian coordinates, the
velocity, and the  Newtonian gravitational potential. The total energy  
density, $E= \rho \epsilon + {1\over  2} \rho v^2$ , is  defined as  the 
sum  of the  thermal  energy, $\rho\epsilon$, where $\rho$ is the 
mass density and $\epsilon$ is  the specific  internal energy,  and  the kinetic
energy  (where  $v^2  =  v_x^2  + v_y^2  +  v_z^2$). Pressure
gradients  and  gravitational  forces  are  the  responsible  for  the
evolution.  An equation of state  $p=p(\rho,\epsilon)$ closes the system. 
We use an ideal gas equation of state $p=(\Gamma-1)\rho\epsilon$ with $\Gamma=2$.


\begin{table*}
  \centering
  \caption{ Overview of the initial models and results of the simulations.
The rows report the name of the model, the ratio of equatorial-to-polar radii 
($r_e/r_p$), the degree of differential rotation ($\hat{A}$), the ratio of kinetic
to potential energy ($T/|W|$), the size of the computational grid ($L$) and the location 
of the corotation radius ($r_c$) for the two resolutions used: high (AMR H) and low (AMR L). 
In models R1H and R2H the corotation radius lies outside the star.
The real (frequency) and imaginary (growth rate) parts of the bar-mode $\sigma_2$ are shown, 
for the low and high resolution simulation in comparison with the numerical results 
and linear analysis by~\cite{shibata02}. Note that for model D3 no linear
analysis results are available.}
\label{tab:initial_models}
\begin{tabular}{llcccccccc}
    \\\hline \hline \\ [-1 em] 
    Model & & D1 & D2 & D3 & R1 & R2 \\ \hline
    $r_e/r_p$ 
    &         & 0.805  & 0.605  & 0.305  & 0.305    & 0.255    \\
    $\hat{A}$    
    &         & 0.3    & 0.3    & 0.3    & 1.0      & 1.0      \\
    $T/|W|$ 
    &         & 0.039  & 0.085  & 0.149  & 0.253    & 0.275    \\
    $L/r_e$ 
    &         & 4.06   & 3.73   & 3.21   & 4.25     & 4.03     \\\hline
    $r_c/r_e$
    & AMR L   & 0.38   & 0.47   & 0.58   &          &          \\
    & AMR H   & 0.36   & 0.48   & 0.56   & -        & -        \\ \hline
    Re$(\sigma_2)/\Omega_0$ 
    & AMR L   & 0.76   & 0.58   & 0.41   &          &          \\
    & AMR H   & 0.81   & 0.55   & 0.43   & -        & 0.82     \\
    & Shibata & 0.80   & 0.60   & 0.45   & 0.92     & 0.75     \\
    & linear  & 0.80   & 0.58   & -      & 0.92     & 0.75     \\\hline
    Im$(\sigma_2)/\Omega_0$ 
    & AMR L   & 0.0042 & 0.0154 & 0.0200 &          &          \\
    & AMR H   & 0.0089 & 0.0190 & 0.0240 & 0.0005   & 0.1960   \\
    & Shibata & 0.009-0.013 & 0.019-0.021 & 0.013  & $<$0.002 & 0.23     \\
    & linear  & 0.015  & 0.021  & -      & $<$0.002 & 0.20     \\
    \hline \hline
  \end{tabular}
\end{table*}

The hydrodynamics equations, Eqs.~(\ref{hydro1}--\ref{hydro3}), can  be
rewritten in flux-conservative form:
\begin{equation}
\frac{\partial  {\bf u}}{\partial t}  + \frac{\partial{\bf  f(\bf u)}}
{\partial   x}   +\frac{\partial{\bf    g(\bf   u)}}{\partial   y}   +
\frac{\partial{\bf h(\bf u)}}{\partial z}= {\bf s(\bf u)}
\label{hypersys}
\end{equation}
\noindent
where ${\bf u}$ is the vector of {\it unknowns} (conserved variables):
\begin{equation}
{\bf u} = [ \rho , \rho v_x,\rho v_y,\rho v_z, E] \ \ .
\end{equation}
\noindent
The  three {\it flux}  functions ${\bf  F}^{\alpha} \equiv  \{{\bf f},
{\bf g},{\bf h}\}$ in the spatial directions $x, y, z$, respectively,
are defined by
\begin{eqnarray}
{\bf f(\bf u)} &=& \left[ {\rho v_x}, {\rho v_x^2}
+ p, {\rho v_xv_y},  {\rho v_xv_z},{(E+p)v_x} \right]
\end{eqnarray}
\begin{eqnarray}
{\bf g(\bf u)}  &=& \left[ {\rho v_y} , {\rho v_xv_y},
{\rho v_y^2}  +{p},{\rho v_yv_z},{(E+p)v_y}\right]
\end{eqnarray}
\begin{eqnarray}
{\bf h(\bf u)}  &=& \left[ {\rho v_z} , {\rho v_xv_z},
{\rho v_yv_z},{\rho v_z^2} + {p},{(E+p)v_z}\right]
\end{eqnarray}
\noindent
and the {\it source terms} ${\bf s}$ are given by
\begin{eqnarray}
{\bf   s(\bf    u)}   &=& \left[   0    ,-{\rho}
\frac{\partial\phi}{\partial   x} ,   
-{\rho}  \frac{\partial\phi}{\partial y} ,
-{\rho} \frac{\partial\phi}{\partial z} , 
{\rho v_x}\frac{\partial\phi}{\partial x} -
{\rho v_y}\frac{\partial\phi}{\partial  y}  
-  {\rho v_z}\frac{\partial\phi}{\partial
z}\right].
\end{eqnarray}
\noindent
System (\ref{hypersys}) is  a three-dimensional hyperbolic system
of  conservation laws with  sources ${\bf  s}({\bf  u})$.  

\section{Numerical approach}
\label{numerics}

For our study of the low $T/|W|$ bar-mode instability we perform high-resolution 
simulations of rotating neutron stars using a Newtonian AMR hydrodynamics code 
called {\tt MASCLET}~\cite{quilis04}. The implementation of the AMR technique 
in the code follows the procedure developed by~\cite{berger89}. The 
hydrodynamics equations are solved using a high-resolution shock-capturing
scheme based upon Roe's Riemann solver and second-order cell reconstruction
procedures, while Poisson's equation for the gravitational field is solved
using multigrid techniques. The accuracy and performance of the {\tt MASCLET}
code has been assessed in a number of tests~\cite{quilis04}. We note that the
code was originally designed for cosmological applications, and here
it is applied to simulations of self-gravitating stellar objects for the
first time.

The simulations are performed with two different 
grid resolutions. The low resolution grid consists of a box of size $L$ with 
$128^3$ zones, yielding a fixed resolution of $L/128$. We note that the effective 
resolution of our coarse grid is comparable to that used by~\cite{shibata02}.
Correspondingly, the high resolution grid consists of a base coarse grid 
of $128^3$ cells, and one level of refinement composed of patches with 
maximum size of $64^3$ cells ($32^3$ coarse cells).  This yields a grid 
resolution on the finest grid of $L/256$. This resolution is enough
to resolve the structures simulated, and hence no deeper refinement levels
are needed.
The patches are dynamically allocated covering those regions 
of the star where the highest resolution is required (highest densities).
Typically only one patch is needed for spheroidal models, and 4-8
in models with toroidal topology. The use of AMR techniques in our
high resolution simulations, allows us to save about a factor 4 in 
CPU time and memory with respect to a unigrid simulation with $256^3$ cells.
No symmetries are imposed in 
the simulations. To the best of our knowledge, in the investigations of the 
bar-mode instability performed by previous groups, grid resolutions as high 
as the ones we use here were never employed.

As customary in grid-based codes \cite{font00, duez03} the vacuum
surrounding the star is filled with a tenuous numerical atmosphere
with density $\rho/\rho_{\rm max}\approx 10^{-12}$ and zero
velocities, $\rho_{\rm max}$ being the maximun density. Every grid
cell with $\rho/\rho_{\rm max} < 10^{-6}$ is reset to the atmosphere
values. A correct treatment of the atmosphere is essential for an
accurate description of the stellar dynamics and correct computation
of the growth rates of unstable modes. We have checked that values for
the atmosphere higher than those we chose or a free evolution of the
atmosphere altogether, lead to remarkable changes in the mode
behaviour, growth rates, and frequencies. We have also checked that
lower values for the atmosphere do not produce those changes, which
ensures that our evolutions are not affected by the atmosphere values
used in the simulations.

\section{Results}
\label{results}

\subsection{Initial data}

Differentially rotating stellar models in equilibrium are built according to
the method of~\cite{eriguchi_84}, and used as initial data for the AMR evolution 
code. The stars obey a polytropic equation of state $P=K\rho^\Gamma$ with index 
$\Gamma=2$. As~\cite{shibata02} the profile of the angular velocity $\Omega$ 
is given by 
\begin{eqnarray}
\Omega = \frac{\Omega_0 \hat{A}^2}{(\varpi/r_e)^2 + \hat{A}^2},
\end{eqnarray}
where $r_e$ is the equatorial radius of the star, $\Omega_0$ is the central 
angular velocity, $\varpi$ is the distance to the rotation axis, and $\hat{A}$ 
parametrizes the degree of differential rotation, from $\hat{A}\ll1$ for highly 
differentially rotating stars to $\hat{A}\to\infty$ for rigidly rotating stars. 
For comparison purposes these parameters are chosen as in some of the models 
of~\cite{shibata02}, and are summarized in Table~\ref{tab:initial_models}. 
Models labelled D rotate with a high degree of differential rotation, as 
$\hat{A}=0.3$, and may therefore be subject to the low $T/|W|$ bar-mode 
instability. We also consider models almost rigidly rotating, labelled R, 
prone to experience the ``classical" bar-mode instability. Labels L and H 
in the models refer to low and high resolution respectively.

Following~\cite{shibata02} we perturb the initial density profile $\rho^{(0)}$ 
according to
\begin{eqnarray}
\rho = \rho^{(0)} \, \left(1 + \delta \frac{x^2-y^2}{r_e^2}\right),
\end{eqnarray}
the perturbation of the pressure given by the equation of state accordingly. A perturbation 
amplitude $\delta=0.1$ is used in all our simulations. As we show below this form of 
the perturbation excites the $l=m=2$ bar-mode. In addition, grid discretization 
can leak small amounts of energy to all other possible modes, which could in 
principle grow provided they were unstable and the simulations were carried 
on for sufficiently long times.

\subsection{Stability analysis}

\begin{figure}[t]
\center
\resizebox{0.8\textwidth}{!}{\includegraphics*{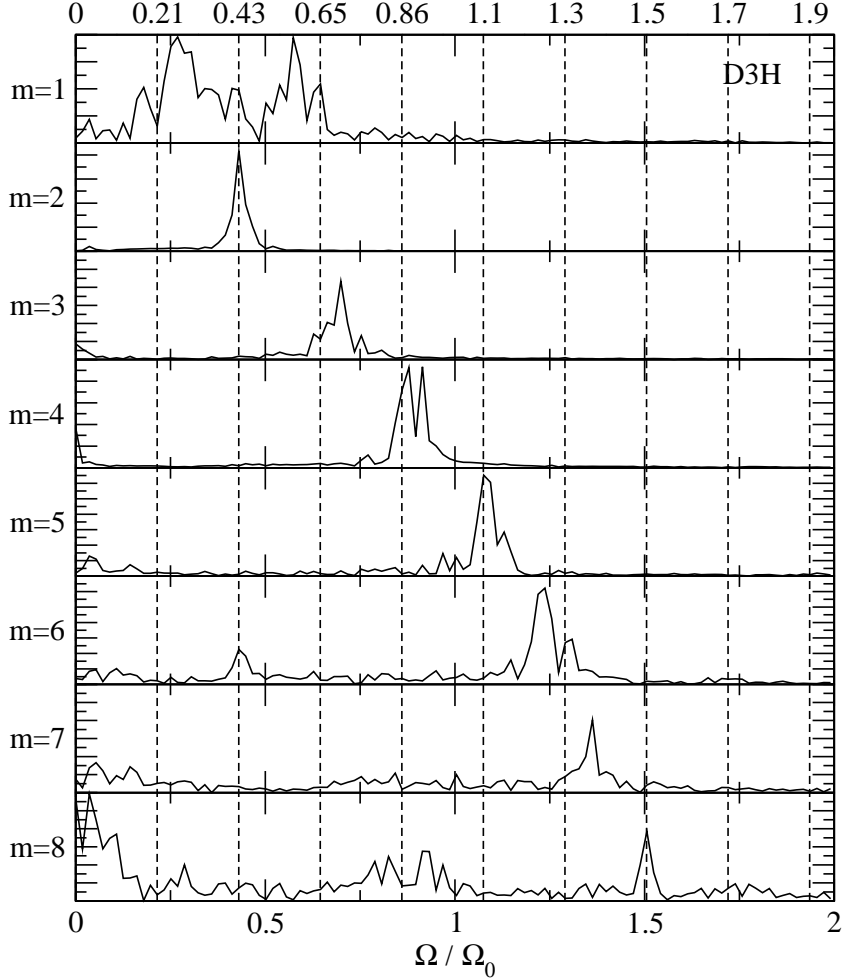}}
\caption{Power spectra of $\mathcal{A}_{m}$ from $m=1$ to $m=8$ for model D3H.}
\label{fig1}
\end{figure}

To compare with~\cite{shibata02} we calculate the distortion parameters
$\eta_{+}$ and $\eta_{\times}$ (and $\eta=(\eta_{+}^{2}+\eta_{\times}^{2})^{1/2}$) 
defined as 
\begin{eqnarray}
\eta_+ \equiv \frac{I_{xx}-I_{yy}}{I_{xx}+I_{yy}}, \hspace{1cm}
\eta_{\times} \equiv \frac{2 I_{xy}}{I_{xx}+I_{yy}},
\label{eq_eta}
\end{eqnarray}
where $I_{ij} (i, j = x, y, z)$ is the mass-quadrupole moment
\begin{eqnarray}
I_{ij} = \int d{\bf x}^3 \rho\, x^i x^j. 
\end{eqnarray}

\begin{figure}[t]
\center
\resizebox{0.8\textwidth}{!}{\includegraphics*{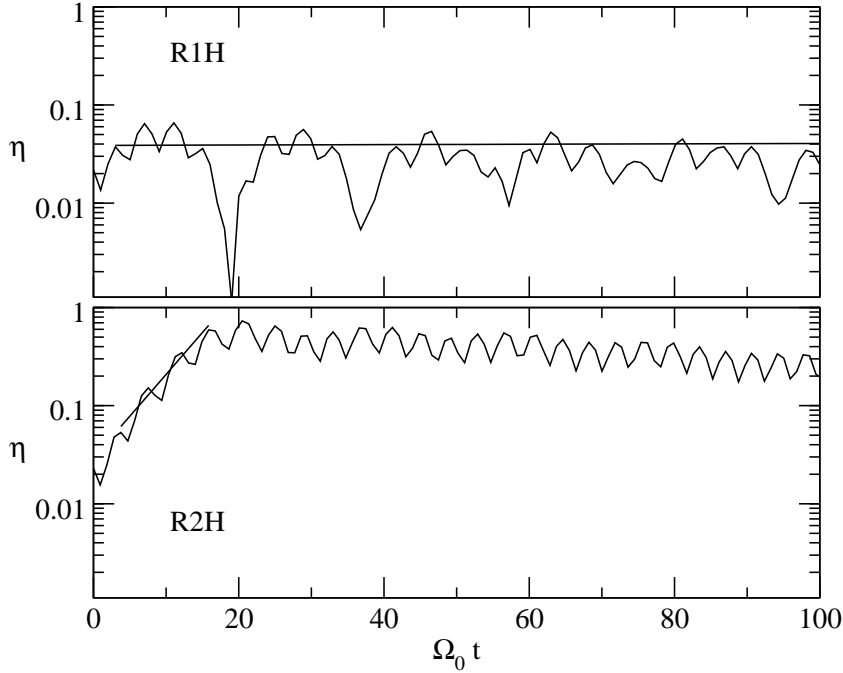}}
\caption{Evolution of $\eta$ for models R1H (upper panel) and R2H (lower panel). 
Exponential fits to the peaks in the growing phase are overplotted as solid lines.}
\label{fig2}
\end{figure}

For the study of the growth rate and interaction of the different angular modes
within the star is useful to calculate the global quantity 
\begin{eqnarray}
A_{m} = \int d{\bf x}^3 \rho ({\bf x}) \, e^{-im\varphi}, 
\end{eqnarray}
and $\mathcal{A}_{m}\equiv 
A_{m}/A_{0}$. 
We follow the time evolution of modes with $m$ ranging from $1$ to $8$.
Since our initial equilibrium models are axisymmetric and have 
equatorial plane symmetry, all $\mathcal{A}_m$ are zero initially, but once perturbed 
all initial models exhibit a dominant $m=2$ component. Assuming that the 
modes behave as $e^{-i(\sigma_m t-m\varphi)}$, the real part of $\sigma_m$ 
can be obtained by Fourier transforming $\mathcal{A}_{m}$. In particular 
Re$(\sigma_2)$, the bar-mode frequency, can be extracted from either 
$\mathcal{A}_2$ or $\eta$ as both represent the same mode. This is the 
dominant mode in all our simulations and its frequency and growth rate 
are given in Table~\ref{tab:initial_models}. The latter corresponds to 
the imaginary part of $\sigma_2$, which is calculated fitting an 
exponential to the peak values of $\eta$ in the growing phase of the 
evolution until the modes saturate. Other modes are also identified 
in the simulations for values of $\mathcal{A}_{m}$ with lower amplitudes. 
We have checked that these modes are harmonics of the $l=m=2$ mode so 
that they follow to good accuracy the relation $\sigma_m = m \sigma_p$, 
$\sigma_p$ being the pattern frequency, calculated as $\sigma_p=\sigma_2 / 2$.
This is shown for model D3H in Fig.~\ref{fig1} which displays the
spectrum of $\mathcal{A}_{m}$ from $m=1$ to $m=8$ (in arbitrary units).
The vertical dashed lines in this figure indicate the location of the
integer multiples of the pattern frequency $\sigma_p$, their values indicated
on the axis at the top of the figure. Each spectrum for each mode is
normalized to its own maximum for plotting purposes. Note that the lower
the mode amplitude the noisier the spectrum and the less accurate the
relation $\sigma_m = m \sigma_p$.

For the models of our sample subject to the ``clasical'' bar-mode deformation (R1H
and R2H), our simulations yield a value of $\beta$ between 0.253 and 0.275, in good
agreement with the critical value for the onset of the dynamical bar-mode instability.
Model R1H is stable and model R2H is unstable. The growth rates and frequencies 
reported in Table~\ref{tab:initial_models} agree with those of~\cite{shibata02}. 
Note that for model R1H, which is stable, the frequency for the $m=2$ mode cannot
be computed. The time evolution of $\eta$ for these two models is displayed in 
Fig.~\ref{fig2}. For the unstable model R2H, our simulations show the formation 
of a bar which saturates for values of $\eta_+$ and $\eta_{\times}$ close to 1, 
i.~e.~in the full nonlinear regime.

\begin{figure}[t]
\center
\resizebox{0.8\textwidth}{!}{\includegraphics*{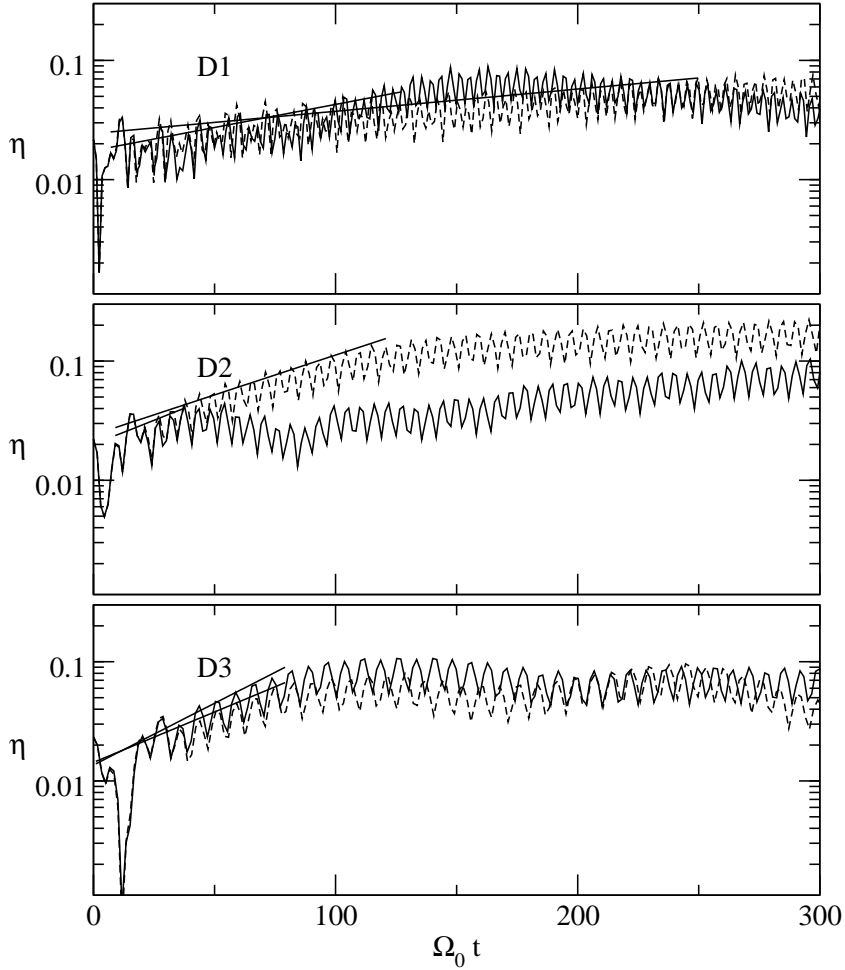}}
\caption{Evolution of $\eta$ for models D1 (upper panel), D2 (central panel) 
and D3 (lower panel). Dashed lines correspond to low resolution and solid lines 
to high resolution. Exponential fits to the peaks in the growing phase are 
overplotted as solid lines.}
\label{fig3}
\end{figure}

Fig.~\ref{fig3} shows the time evolution of $\eta$ for models D in our sample,
prone to suffer the low $T/|W|$ bar-mode instability. Solid lines correspond to
high resolution simulations and dashed lines to low resolution. For all three 
models the pattern frequencies $\sigma_p$ are such that there exists a corotation 
radius inside the star, i.e.~a radius at which the bar-mode rotates with the same 
angular velocity as the fluid. The location of the corotation radius for all models
of our sample is reported in Table~\ref{tab:initial_models}. As recently discussed 
by~\cite{watts05} the existence of such corotation radius is a potential 
requirement for the ocurrence of the instability. As becomes clear from 
Fig.~\ref{fig3}, all models are unstable but grid resolution has an important 
effect on the saturation of the instability once the nonlinear phase has been 
reached, as well as in the long-term dynamics of the stars. 

In the linear phase of models D1H and D2H, the growth rates and frequencies 
agree with the results of~\cite{shibata02} in both, the numerical simulations
and the linear analysis (see Table~\ref{tab:initial_models}). In the linear phase
of model D3H, our frequencies are similar to the numerical results 
of~\cite{shibata02}, although our growth rates are about a factor two larger.
We emphasize that no results are reported in the linear analysis for this model
in the work of~\cite{shibata02}, and therefore this discrepancy can be 
an effect of the resolution used or of the characteristics of each numerical
code. Increasing resolution leads to similar results in 
the frequencies but to higher growth rates.

In the nonlinear phase, models D1 and D3 
behave similarly for the two resolutions used (see Fig.~\ref{fig3}),
and also similarly to the results by~\cite{shibata02} (compare with Fig.~3
of that paper). For model D2 we observe a radical change of behavior in the 
nonlinear phase of the mode evolution depending on the grid resolution. This has 
implications on the long-term dynamics of the star and, in particular, on the 
attainable amplitudes of the gravitational radiation emitted, as we discuss below. 

\begin{figure}[t]
\center
\resizebox{0.8\textwidth}{!}{
\includegraphics*{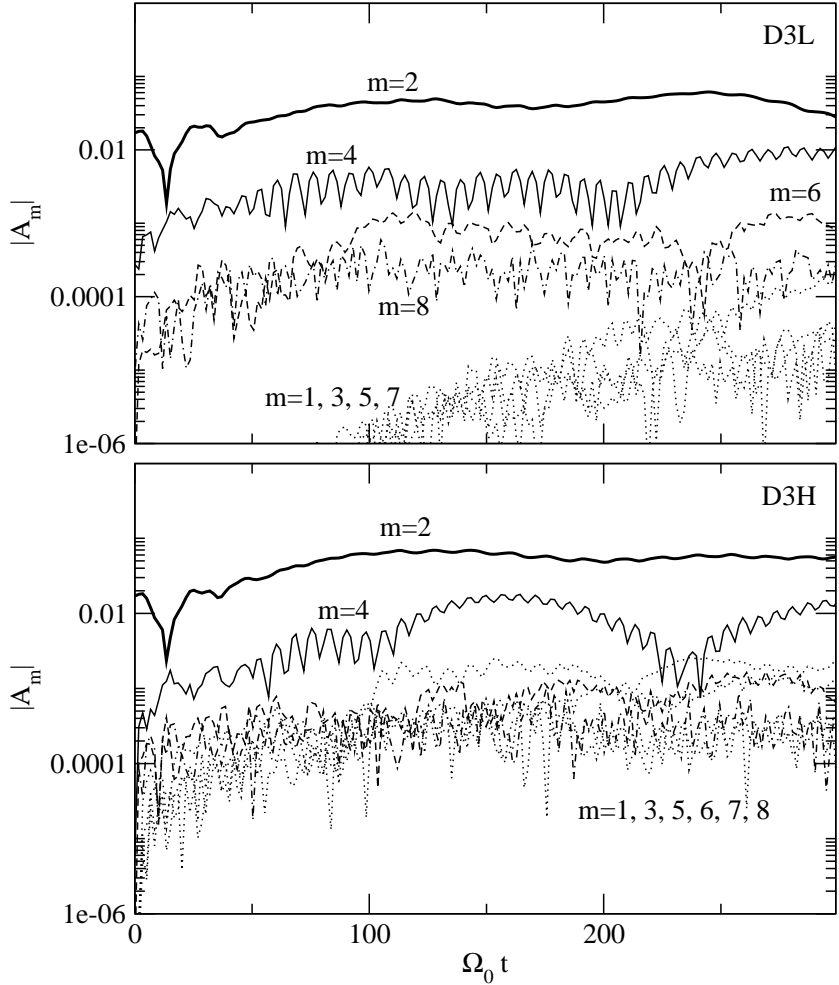}}
\caption{Evolution of $|\mathcal{A}_{m}|$ for model D3 with low resolution (top)
and high resolution (bottom). The $m=2$ mode is represented with thick solid line,
$m=4$ with thin solid line, $m=6$ with dashed line, $m=8$ with dot-dashed line,
and all other odd $m$ with dotted lines.}
\label{fig4}
\end{figure}

It is worth mentioning the possibility that the unstable mode at the start of
model D2H might excite some other mode in the corotation band, which could not 
otherwise be excited for lower grid resolution. As discussed 
by~\cite{watts03,watts04} in their study of differentially rotating shells, there
are many zero-step modes in the band, so that the whole continuous spectrum could
potentially be excited. In such case these modes would have very slow power-law
growth.

For all our models we have checked mass conservation along the evolution. 
The worst results are obtained for model D3H, for which mass is conserved
within $2.5\%$ error when the instability saturates. At the end 
of the simulation (after 48 orbital periods and 25000 iterations in the 
coarsest grid) the error has grown to only $6\%$. For all other models mass 
conservation is even more accurate.
Note that these errors are within the round-off error of the code, and it is not 
related to the conservation properties of the numerical scheme itself. For a
regular grid with $128^3$ cells and a simulation employing $25000$ iterations, 
the accumulated round-off error
(binomial distribution) using single-precision arithmetics, is about 
$\sqrt{128^3 \times 25000} \times 10^{-8} = 0.0023 = 0.23 \%$. Correspondingly, for a $256^3$ grid 
(with twice the number of iterations for the simulation) the error is about $0.9\%$. Taking into account that this
error affects the nonlinear evolution of the system, it is not surprising
to have an error at the level of a few percent by the end of our high resolution simulations, 
for all conserved quantities.

Figure \ref{fig4} shows the evolution of $\mathcal{A}_{m}$ for model D3 and 
for $m$ ranging from 1 to 8 for our two resolutions. According to this figure,
the only two modes relevant for the dynamics of the star are $m=2$ and $m=4$. 
All other modes have smaller amplitudes and play no role in the dynamics. Note 
that for odd $m$ modes, the value of the integrated quantity $\mathcal{A}_{m}$, 
if close to zero, is extremely sensitive to very small numerical asymmetries, 
which are induced by the patch creation scheme of our AMR code. This explains 
the resolution differences in the initial values for odd $m$ modes in 
Fig.~\ref{fig4} (at $t=0$ they start off at $10^{-8}$ level for the low
resolution simulation), although they saturate at the same value irrespective 
of the resolution.

\begin{figure}[t]
\center
\resizebox{0.8\textwidth}{!}{
\includegraphics*{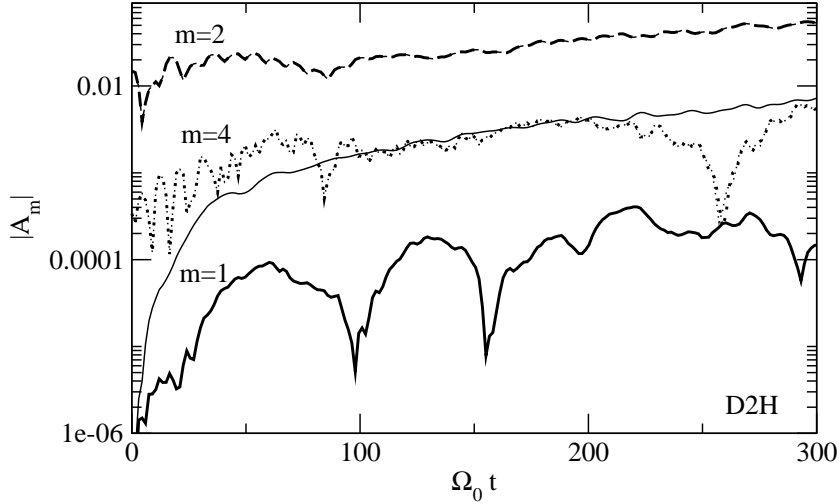}}
\caption{Effects of the artificial displacement of the center of mass (of only 
one numerical cell) on the time evolution of $|\mathcal{A}_{1}|$ for model D2H. The 
thin solid line shows a fictitiuos evolution resulting from the numerical artifact 
originated by the center of mass displacement.}
\label{fig5}
\end{figure}

An important diagnosis for the accuracy of the results is the location
of the center of mass during an evolution. {The round-off error of the numerical code}
imposes controlled errors in mass and linear momentum, which results
in tiny displacements of the center of mass. However small (one numerical
cell in our runs) this unphysical displacement may hinder the correct 
analysis of the mode growth rates. For this reason all integrated quantities 
shown in Fig.~\ref{fig4} are computed after correcting for the displacement 
of the center of mass, ${\bf x}_{\rm new}={\bf x}_{\rm old}-{\bf x}_{\rm CM}$, 
in a post-processing stage of the data analysis. Were this not done, a one-armed $m=1$ 
mode would grow much faster than it should to bring up fictitious features 
in the plots. This is shown for model D2H in Fig.~\ref{fig5}. The thick 
solid line in this figure corresponds to the evolution of the $m=1$ mode 
taking into account the correction for the center of mass displacement, while 
the thin solid line is the corresponding evolution of this mode without the correction.

\subsection{Gravitational waves}

\begin{figure}[t]
\center
\resizebox{0.8\textwidth}{!}{
\includegraphics*{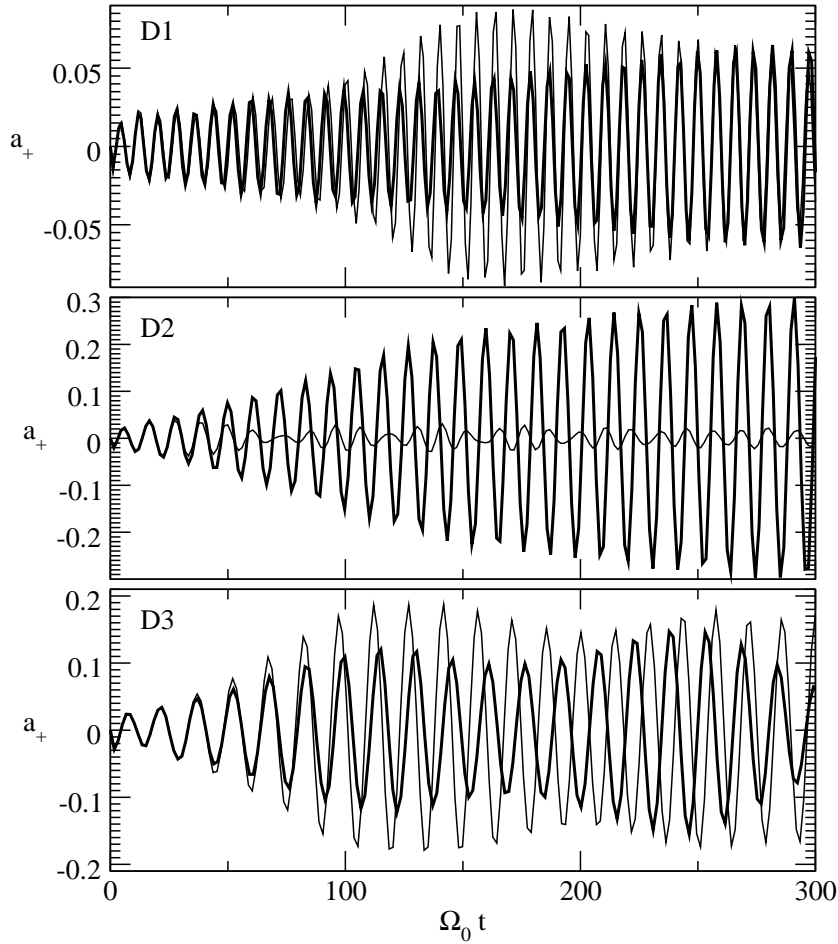}}
\caption{Gravitational waves for models D1 to D3 extracted using the standard
quadrupole formula. Thick (thin) solid lines correspond to low (high) resolution. 
Only the dimensionless waveform amplitude $a_{+}$ is plotted.}
\label{fig6}
\end{figure}

The growth and saturation of the instability is also imprinted on the gravitational
waves emitted. The gravitational waveforms $h_{+}$ and $h_{\times}$ for  models
D1, D2, and D3, computed using the standard quadrupole formula, are shown in  
Fig.~\ref{fig6}. For a source of mass $M$ located at a distance $R$ 
those waveforms can be calculated from the dimensionless waveform 
amplitudes $a_{+}$ and $a_{\times}$ as
\begin{eqnarray}
h_{+,\times} = a_{+,\times} \frac{\sin^2{\theta}}{R} \frac{M^2}{r_e}, 
\end{eqnarray}
using $G=c=1$ units. The resulting chirp-like signal in all the models, particularly 
apparent for model D2L, indicates the presence of a bipolar distribution of mass 
within the star (see Sec.~\ref{morpho}).

As mentioned before, the effects of grid resolution on the evolution of the 
nonlinear phase of the bar-mode are imprinted on the gravitational waveforms. 
Thick solid lines in Fig.~\ref{fig6} are the waveforms which correspond to the 
low-resolution models, and thin solid lines to the high-resolution counterparts. 
The evolution of $\eta$ for model D3, displayed in Fig.~\ref{fig3}, shows little 
deviations with grid resolution, and this translates into very similar gravitational 
wave patterns (bottom panel of Fig.~\ref{fig6}), the differences becoming more 
noticeable in the nonlinear phase following saturation ($\Omega_0 t \geq 75$).
For model D1 (top panel), the differences also become more apparent at later times
during the evolution, in good agreement with the dissimilar behaviour of the matter 
dynamics in this model, as encoded in the evolution of $\eta$ in Fig.~\ref{fig3}. 
As happens for model D3 the first few cycles of the gravitational waveform, when 
the mode is still in the linear phase, are accurately captured for both resolutions.

\begin{figure*}[t]
\begin{center}
\resizebox{\textwidth}{!}{\includegraphics*{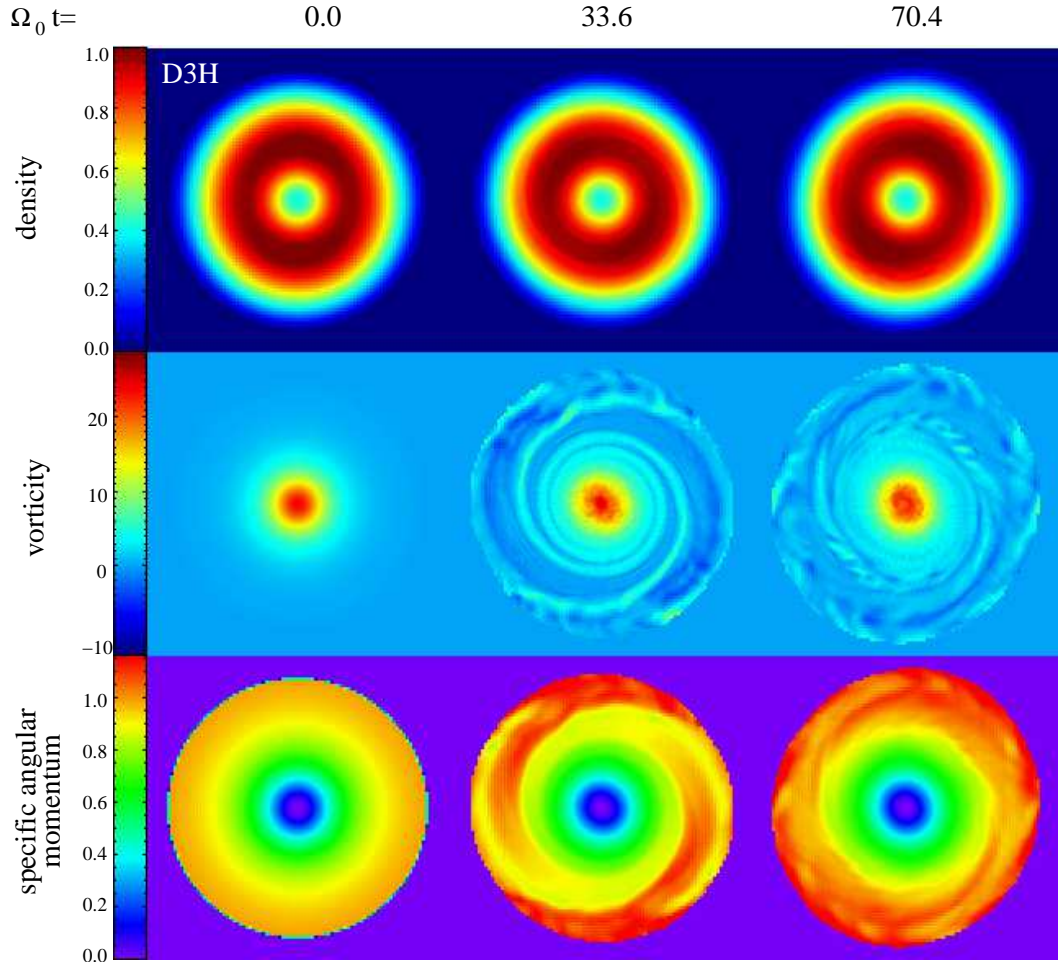}}
\caption{
Snapshots of the density, vorticity, and specific angular momentum, for model 
D3H, at three representative instants of the evolution. All snapshots show 
slices of the stars in the equatorial plane. Quantities are normalized as follows: 
$\rho/\rho_{\rm max}^{(0)}$, $r_e w^{\varphi}/v_s^{(0)}$, and $l^{\varphi}/
(r_e v_s^{(0)})$, where $v_s^{(0)}$ is the initial velocity at the surface
of the star.}
\label{fig7}
\end{center}
\end{figure*}

The major dependence of the waveform on the grid resolution is found for model D2. 
Again, the linear phase for the growth of the bar deformation is accurately captured 
irrespective of the resolution (and agrees with the perturbative results 
of \cite{shibata02}). This is signalled in the perfect overlapping of both 
gravitational waveforms during the first three cycles (see the middle panel of 
Fig.~\ref{fig6}). However, the different nonlinear dynamics of the bar-mode deformation 
for this model, shown in the middle panel of Fig.~\ref{fig3}, is severely imprinted 
on the gravitational waveform. Model D2H emits gravitational waves which have roughly 
one order of magnitude smaller amplitude than those computed for the corresponding 
low resolution model.

\subsection{Morphology}
\label{morpho}

We next describe the morphological features encountered during the
evolution of some representative models. Fig.~\ref{fig7} shows three 
snaphsots of the evolution of model D3H for the density (top), the azimuthal 
component of the vorticity, $\vec{w}^{\varphi}=(\nabla\times \vec{v})^{\varphi}$ 
(middle), and the specific angular momentum, $\vec{l}=\vec{r}\times\vec{v}$ 
(bottom). From left to right the snapshots correspond to the initial time 
($\Omega_0 t=0$), a time when the bar-mode instability is growing 
($\Omega_0 t=33.6$), and the time when the instability saturates ($\Omega_0 
t=70.4$). Only the equatorial plane of the stars is shown in all these plots. 
Animations of all simulations performed are available at 
{\tt www.uv.es/$\sim$cerdupa/bars/}. We note that our AMR code is able to 
dynamically place patches (e.~g.~between 4 and 8 in the D3H model) and evolves
the system with continuous matching between patches, as exemplified in 
Fig.~\ref{fig7}.

The evolution of model D3H shows that as the $m=2$ mode grows the star develops 
an ellipsoidal shape which remains spinning beyond saturation. Since the low 
$\beta$ $m=2$ mode saturates at lower values ($\eta\sim0.1$) than the 
classical bar-mode instability ($\eta\sim1$), no clear bars are visible 
in the density plot. At late times ($\Omega_0 t > 100$) a ``boxy'' structure 
becomes apparent as the $m=4$ mode has grown to almost similar amplitude as the $m=2$ mode (see 
animations and Fig.~\ref{fig4}). No other global features can be seen, consistent 
with the fact that $|\mathcal{A}_m|\ll1$ for all modes other than $m=2$ and 4. The 
vorticity plot shows that the $m=2$ mode at $\Omega_0 t=33.6$ adopts the form of 
a two-armed spiral winding up around the central parts of the star. As the mode 
begins to saturate ($\Omega_0 t=70.4$) the spirals break apart into the outer layers in 
a turbulent flow reminiscent of the (shear) Kelvin-Helmholtz instability, and 
shock as they reach the atmosphere. These trends are also 
visible in the specific angular momentum plot. 

\begin{figure}[t]
\begin{center}
\resizebox{\textwidth}{!}{\includegraphics*{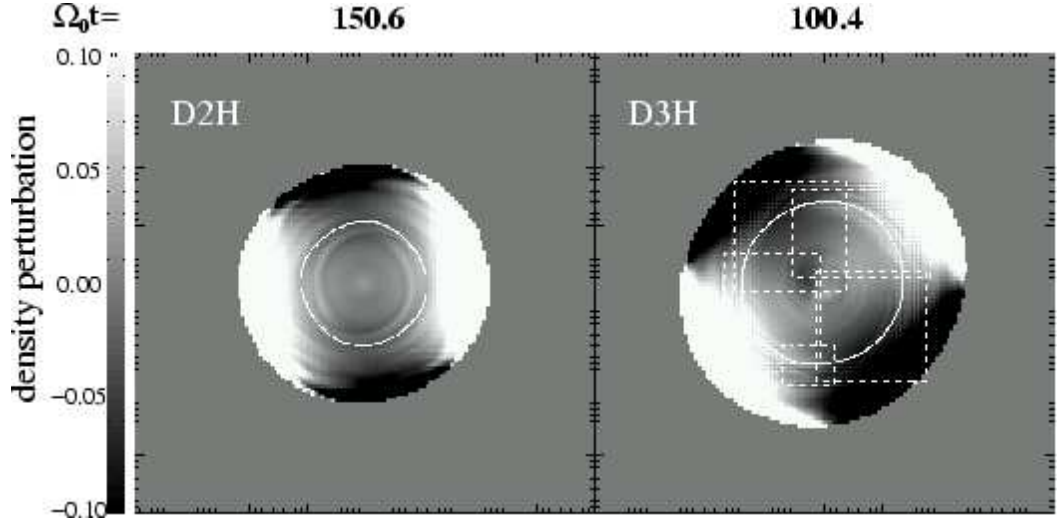}}
\caption{Snapshots of the density perturbation at the equatorial plane
for models D2H and D3H. The white solid curves indicate the location
of the corotation radius. The white dashed boxes indicate the location 
of the patches for model D3H.}
\label{fig8}
\end{center}
\end{figure}

The presence of a corotation radius, at $r/r_e=0.56$ for model D3H, seems to
play a role in the growth and saturation of the instability, in agreement with
the recent findings of~\cite{saijo05b}. As the bar-mode grows, pressure waves 
carry angular momentum outside the corotation radius, which is deposited
in the outer layers of the star. This excites Kelvin-Helmholtz-like instabilities 
in the fluid that break the mode outside the corotation radius. When this happens the 
$m=2$ instability stops growing and no more angular momentum is extracted. 
Figure~\ref{fig8} shows late-time snapshots of the equatorial plane distribution 
of the density perturbation, i.e.~$(\rho-\rho^{(0)})/\rho^{(0)}_{\rm max}$, for 
models D2H and D3H. The times are chosen well inside the nonlinear and saturation 
phase of the instability. This figure helps to interpret the mode dynamics and its 
saturation along the lines mentioned before: During the evolution the density 
perturbations are shed in waves from the center towards the outer layers of the 
star. At late times, when the instability saturates, such shedding stops, and the 
density perturbation reaches the largest values outside the corotation radius 
(depicted with white solid lines in Fig.~\ref{fig8}), for either model.

We note in passing that the corotation radius in all our high resolution models 
lies well inside the outer boundary of the finest box set up by the AMR refinement 
pattern. (see, e.g. the white dashed boxes depicted in the right panel of 
Fig.~\ref{fig8} indicating the location of the AMR patches for model D3H)
This rules out the possibility of a numerical artifact resulting from the 
patch creation scheme of our AMR code being the cause for the different long-term 
evolution between low and high resolution models, particularly noticeable for
model D2 in Fig.~\ref{fig3}.

\begin{figure}[t]
\begin{center}
\resizebox{\textwidth}{!}{\includegraphics*{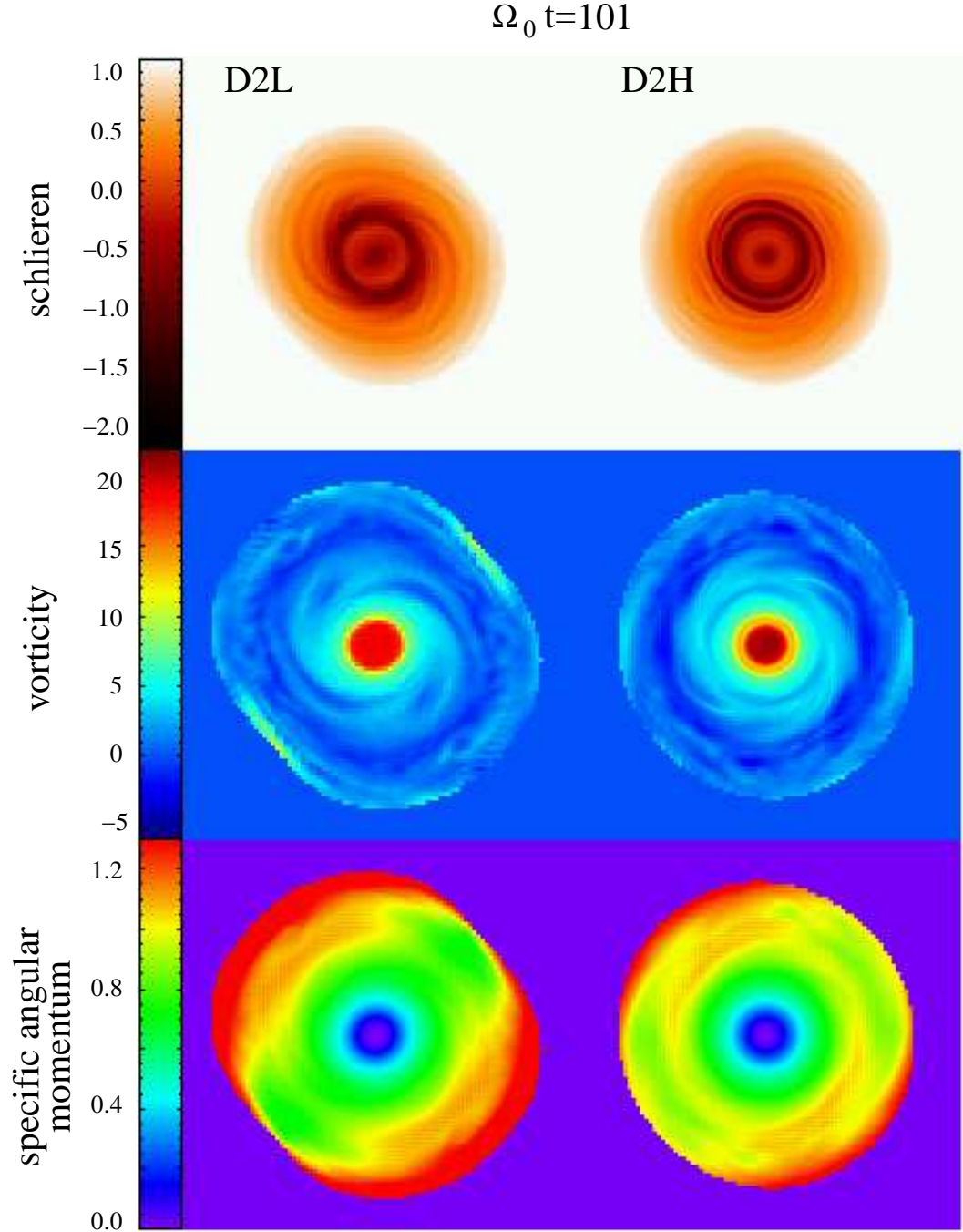}}
\caption{
Resolution comparison between models D2L and D2H once the instability has saturated. 
Only slices of the stars in the equatorial plane are shown.}
\label{fig9}
\end{center}
\end{figure}

Finally, Fig.~\ref{fig9} shows a comparison between models D2L and D2H at 
$\Omega_0 t=101$ (i.e.~well within the nonlinear phase), to highlight the 
effects of the numerical resolution on the morphology. From top to bottom 
this panel shows a schlieren plot ($|\nabla\log\rho|$) , $\vec{w}^{\varphi}$, 
and $\vec{l}$. The resolution differences in the evolution of model D2 become 
apparent from this figure. In particular, the ``boxy" structure becomes much 
more clearly visible in the low resolution simulation (D2L), indicating an 
excessive growth rate of the $m=4$ mode. The presence of pressure waves is 
emphasized in the schlieren plot, very accurately captured in model D2H.
Those waves, once the flow is driven to turbulence past the corotation radius,
redistribute the angular momentum in the outer layers of model D2L in a much more
pronounced way than for model D2H.

\section{Summary and outlook}
\label{conclusions}

We have presented AMR high-resolution simulations of the low $T/|W|$ bar-mode 
instability of extremely differentially rotating neutron stars. Our main 
motivation has been to revisit the simulations by~\cite{shibata02} on such 
instability, assessing how sensitive the onset and development of the instability
is to numerical issues such as grid resolution. We have addressed the importance 
of a correct treatment of delicate numerical aspects which may spoil 
three-dimensional simulations in (Cartesian) grid-based codes, always hampered 
by insufficient resolution, namely the handling of the low-density atmosphere
surrounding the star, the correction for the center of mass displacement, and the
mass and momentum conservation properties of the numerical scheme. Our simulations 
have revealed the complex morphological features involved in the nonlinear 
dynamics of the instability. We have found that in the nonlinear phase of
the evolution, the excitation of Kelvin-Helmholtz-like fluid modes outside 
the corotation radii of the stellar models leads to the saturation of the bar-mode 
deformation. While the overall trends reported in the investigation 
of~\cite{shibata02} are confirmed by our work, the resolution used to perform the  
simulations may play a key role on the long-term behaviour of the instability 
and on the nonlinear dynamics of rotating stars, which has only become apparent 
for some specific models of our sample (namely model D2). This, in turn, has 
implications on the attainable amplitudes of the associated gravitational wave 
signals.

The work reported in this paper is a first step in our ongoing efforts of 
studying the dynamical bar-mode instability within the magnetized core collapse 
scenario.

\section*{Acknowledgements}

The authors thank Harry Dimmelmeier, Nick Stergioulas, and Anna Wats for useful 
comments. Research supported by the  Spanish {\it  Ministerio de  Educaci\'on y
Ciencia} (MEC; grants  AYA2004-08067-C03-01,   AYA2003-08739-C02-02, AYA2006-02570).
VQ  is a Ram\'on  y Cajal Fellow  of the  Spanish MEC. Computations  performed  
at  the  {\it Servei  d'Inform\'atica  de  la Universitat de Val\`encia} 
({\tt CERCA-CESAR}).


\end{document}